\documentclass[jeosman,preprint,fleqn,showpacs,showkeys]{revtex4}
\usepackage{graphicx}
\usepackage{amssymb,amsmath,bm}

\begin{document}

\title{Engineering sub-Poisson light in a simple mirror and beam splitter system}

\author{Sun-Hyun Youn\footnote{E-mail: sunyoun@jnu.ac.kr, fax: +82-62-530-3369}}
\address{Department of Physics, Chonnam National University, Gwangju 500-757, Korea}

\begin{abstract}

Vacuum fluctuation, which is the intrinsic nature of an electric field
 can be measured via homodyne detection. Moreover, electric field intensity fluctuation are also related to vacuum fluctuations.
 Squeezed vacuum and sub-Poisson light can be obtained by controlling the vacuum fluctuation using noble nonlinear interaction. 
Based on the squeezed vacuum by inserting  a mirror on the unused part of the beam splitter was proposed in 1994, we present the mode matching method for the vacuum and light fields. Light intensity fluctuations also can be reduced by inserting a mirror on the unused part of the beam splitter. To obtain sub-Poisson light as a function of the distance between the mirror and detector, a detector with a thinner active layer than the wavelength is required.



\end{abstract}


\maketitle

\section{Introduction}

When a single photon is in a particular mode, according to the particle nature of light, photons will be sequentially found in that mode. The probability of finding a photon is proportional to the absolute square of the wave function related to the electromagnetic wave.
Vacuum fluctuations are related to the spatial characteristics of the electromagnetic wave. The spontaneous decay caused by the vacuum can be suppressed in cavities \cite{Jhe1987}. Theoretical and experimental studies have beem conducted on methods to change the vacuum fluctuations near mirrors\cite{sun1995,sun1994,wadood}.

In this study, in contrast to previous studies on the vacuum noise characteristics of light using a homodyne detector, we calculate the intensity fluctuations when photons are directly measured using photon counter. The obtained results are similar to those obtained in previous studies, but herein we predict the results considering mode matching  in the experiment.

In section II, the fluctuation of light that can be measured using a detector is calculated with a mirror placed on one side of the beam splitter. In section III, an experimental device is proposed for perfect mode matching, and in the last section, the practical limits of the vacuum fluctuation near the mirror are discussed.

\section{Vacuum fluctuation near a mirror.}

\begin{figure}[htbp]
\centering
\includegraphics[width=8cm]{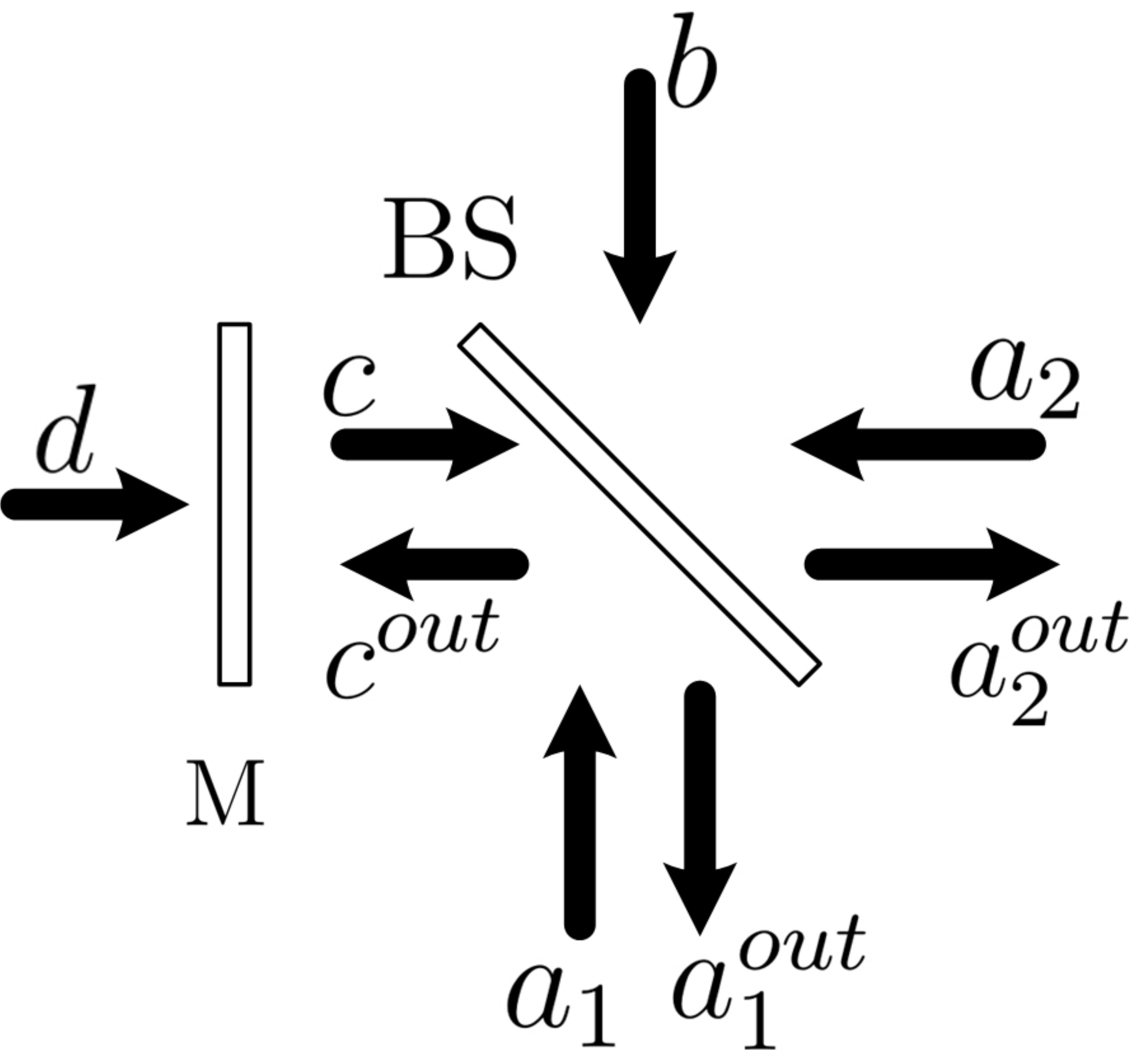}
\caption{Vacuum mode relations in the beam splitter with a mirror.  BS: Beam splitter, M: mirror}
 \label{DoubleP}
\end{figure}
An electric field can be written as
\begin{eqnarray}
\hat{E}_{L} = \hat{E}_{cl} + \hat{E}_{Q} ,
  \label{eLo}
\end{eqnarray}
where
\begin{eqnarray}
\hat{ E} _{cl}  &=&   i \sqrt{\frac{\hbar \omega }{2 \epsilon_0 V }} ( \alpha  e^{i(\omega t  - k_0 z )}  - \alpha^{*} e^{i(\omega t - k_0 z )} ) \vec{x},  \nonumber \\ 
\hat{ E} _{Q}   &=& i \sum_{k}  \sqrt{\frac{\hbar \omega_k }{2 \epsilon_0 V }} ( \hat{b}_{k} e^{- i (\omega_k  t - k z)   }
-\hat{b}_{k}^{\dagger}  e^{ i (\omega_k  t - k z)   } )\vec{x} .
  \label{efcleq}
\end{eqnarray}
Here, $k_0$ and  $\omega$ are the wave number  and angular frequency of the laser, respectively,  $\hbar$ and $\epsilon_0$ have usual meanings, and $V$ is 
the normalization volume\cite{yariv}.
Considering the laser mode in Fig. \ref{DoubleP}, the modes $a_1 ^{out} $ and $ a_2 ^{out} $ can be written as
\begin{eqnarray}
a_1 ^{out} &=&   \sqrt{T} b  + \sqrt{R} c  ,  \nonumber \\ 
a_2 ^{out} &=&  -\sqrt{R} b  + \sqrt{T} c ,
\label{a1a2out} 
\end{eqnarray}

where the modes $c$ and $c^{out} $ can be written as
 \begin{eqnarray}
c  &=& \sqrt{T_m} d  - \sqrt{R_m} c^{out} , \nonumber \\
 c^{out} &=& \sqrt{ R }a_1 + \sqrt{T} a_2 .\nonumber 
\label{cmode} 
\end{eqnarray}
Then  the electric field in fluctuating vacuum modes at $a_1$ is 
\begin{eqnarray}
\hat{ E} _{vac,1} ^{(+)}  &=& \sum_{k} i \sqrt{\frac{\hbar \omega_k }{4 \epsilon_0 V }} \{ \sqrt{T} \hat{b}_{k}^{\dagger} e^{i(\omega_k t  - k Z_1)} + \mu  \hat{a}_{1,k}^{\dagger} e^{i(\omega_k  t + k z_1 )} \nonumber \\ 
& &-  R \sqrt{R_m}  \hat{a}_{1,k}^{\dagger}  e^{i(\omega_k t  - k z_1 )} - \sqrt{RT } \sqrt{R_m} \hat{a}_{2,k}^{\dagger}  e^{i(\omega_k t - k z_1) } + \sqrt{ R T_m} \hat{d}_{k} ^{\dagger} e^{i(\omega_k t - k Z_M )} \}
  \label{efield1}
\end{eqnarray}
where $R_m$($T_m$) is the reflectance(transmittance) of the mirror and $R$($T$) is the  reflectance (transmittance) of the beam splitter, $z_1$($Z_1$) is the distance from the mirror (laser) to the detector. $Z_M$ is related to the vacuum source behind the mirror and it can be any number.
  We add the factor $\frac{1}{\sqrt{2}}$ for the normalization of the vacuum fluctuation.  The vacuum mode ($\hat a_1 ^{\dagger} e^{i ( \omega t - k z_1 )}$)  at the detector is the reflected vacuum mode  ($\hat a_1 ^{\dagger} e^{i ( \omega t + k z_1 )}$)  at the mirror. If two modes are perfectly matched the $\mu$ in Eq. \ref{efield1} is 1 and the two counterpropagating modes yield the standing wave mode\cite{sun1994,sun1995}.  
If $\mu = 0 $, the fluctuation value from Eq. \ref{fluc} becomes $\frac{|\alpha|^2 T}{2}$, it is the square of the constant dc current $ \frac{T |\alpha |^2}{2} $.
In other words, if we directly measure the fluctuation of the laser intensity, the fluctuation is dependent on the distance ($z_1$) between the mirror and the detector.
Even in photo counting experiments, the photon number fluctuation is  related to the vacuum fluctuation, therefor, the photon number fluctuation is also
depend on the distance $z_1$.

If we used the photodetetion theory \cite{detection} with instantaneous response of the photodetector  \cite{response}, 
\begin{eqnarray}
 \hat{I}_1 = \{ \sqrt{T} \hat{E}_{cl}^{(+)} + \hat{E}_{vac,1}^{(+)} \}  \times \{ \sqrt{T} \hat{E}_{cl}^{(-)} + \hat{E}_{vac,1}^{(-)}\},
  \label{current}
\end{eqnarray}
where we normalize the photocurrent. If the electric field of the local oscillator is considerably  greater than the vacuum field,  the terms containig  $\alpha$ have physical significance.  When the constant dc current $ \frac{T |\alpha |^2}{2} $ is neglected, Eq. \ref{current} yields 
 \begin{eqnarray}
&& \hat {I}_{1}^{o} (z_1, Z_1)   = \frac{|\alpha|}{\sqrt{2}} [\sqrt{T} e^{i \phi} \{  (\mu  e^{- i k (Z_1 + z_1 )} - e^{- i k (Z_1 - z_1 )} R \sqrt{R_m} ) \hat{a}_1   - e^{- i k (Z_1 - z_1 )} \hat{a}_2  \} \nonumber \\
&+& \sqrt{T} e^{- i \phi } \{  (\mu  e^{ i k (Z_1 + z_1 )} -  e^ { i k ( Z_1 - z_1 )} R \sqrt{R_m} ) \hat{a}_1 ^{\dagger}    - e^{-  k (Z_1 - z_1 )} \hat{a}_2 ^{\dagger}  \} \nonumber \\
&+& e^{ i \phi} T \hat{b} +e^{- i \phi} T \hat{b}^{\dagger} +  e^{i \phi} e^{i k (Z_M - Z_1 )} \sqrt{T R T_m } \hat{d} + 
e^{-i \phi} e^{- i k (Z_M - z_1 )} \sqrt{T R T_m } \hat{d}^{\dagger} ],
\label{photocurrentA} 
\end{eqnarray}

We then  evaluate the square of the photocurrent to determine the fluctuation. After squaring Eq. \ref{photocurrentA}, we find the photocurrent fluctuation as follows:
\begin{eqnarray}
\langle (\hat {I}_1 ^{o})^2  \rangle  =\frac{|\alpha|^2 T}{2}   \{ 1+\mu^2 - 2 \mu R \sqrt{R_m} \cos( 2 k z_1 )\} 
\label{fluc} 
\end{eqnarray}

If $\mu = 0 $, the fluctuation value from Eq. \ref{fluc} becomes $\frac{|\alpha|^2 T}{2}$, which is the square of the constant dc current $ \frac{\sqrt{T} |\alpha |}{\sqrt{2}} $.
In other words, if we directly measure  the laser intensity fluctuation, the fluctuation is dependent on the distance ($z_1$) between the mirror and detector.
Even in the photo counting experiment, the photon number fluctuation is  related to the vacuum fluctuation; therefore, the photon number fluctuation is also
dependent  on the distance $z_1$.
  
If we consider practical limits such as finite linewidth and finite absorption length, Eq. \ref{fluc} will  change as follows\cite{ref7, sun1995}. 

\begin{eqnarray}
\langle (\hat {I}_1 ^{o})^2  \rangle_{P}   &=&\frac{|\alpha|^2 T}{2}   \{ 1+\mu^2 - 2 \mu R \sqrt{R_m} e^{- z_1 ^2 \Delta k^2} \nonumber \\
&\times& \frac{\kappa [ \cos(2 k_0 z_1 + \phi_0 ) - e^{-\kappa D } \cos(2 k_0 (z_1 + D) + \phi_0 )]}{\sqrt{4 k_0 ^2 + \kappa ^2}}  \}
\label{flucPr}, 
\end{eqnarray}
where $\Delta k$ is the line width of the local oscillator beam with Gaussian line width distribution functions. $\kappa$ is the absorption coefficient, $D$ is the detector active length,   and $\phi_0 = \arctan \frac{2k}{\kappa} $.  We assumed that the probability that a photon is converted into an electron hole pair at distance $\eta$ from the surface of the detector's active region is $\kappa e^{- \kappa \eta} $\cite{ref8}.

The two coefficients $\sqrt{R_m}$ and $\mu$ depend on the mode matching condition. Even when we used the total mirror, if the mode from the mirror is not perfectly matched with the mode from the laser, the effective reflectance $ \sqrt{R_m} $ can not be $1$. Furthermore, the mode $a_1$  to the mirror is reflected by the mirror and then meets at the detector. At the detector, if two counter-propagating modes are not exactly matched, the coefficient $\mu$ cannot be $1$. To evaluate this mode matching condition, 
we assume that the amplitude envelope of the electromagnetic wave in the transverse plane is given by a Gaussian function. 

Considering the Gaussian modes \cite{saleh}
\begin{eqnarray}
E(\rho, z ) = E_0 \frac{w_0}{w(z)} \exp [- \frac{\rho^2}{w(z)^2} ] \exp [ - i k z - i k \frac{\rho^2}{2 R(z)} + i \zeta(z)]
\label{gfield}
\end{eqnarray}
, where  $w_0$ is the radius of the beam waist and
\begin{eqnarray}
w(z) &=& w_0 \sqrt{1+ (\frac{z}{z_0})^2} \nonumber \\
R(z) &=& z (1+ (\frac{z_0}{z})^2 ) \nonumber \\
\zeta(z) &=& \tan^{-1} \frac{z}{z_0} 
\label{gfieldsub}
\end{eqnarray}
and $z_0$ is defined as follows:
\begin{eqnarray}
z_0 &=& \frac{\pi}{\lambda } w_0 ^2.  
\label{z0}
\end{eqnarray}

First, we assume that the laser and vacuu modes have the same beam waist $w_0$ at the detector. Then the laser and vacuum modes are perfectly matched; thus,  $\sqrt{R_m} = 1$.  On the other hand, the vacuum  $E_v (0) $ starting from the detector propagates to the mirror and reflects at the mirror. The returned vacuum $E_v (2 z_1 )$ is not the same 
 $E_v(0)$. The coefficient $\mu$ can be calculated as follow:

\begin{eqnarray}
\mu &=&\frac{|<E_v (0)  E_v (2 z_1 )^{*} >| }{\sqrt{<E_v (0)^2 > <E_v (2 z_1 )^2>}}
  \nonumber \\
  &=&  \frac{(1+ \frac{4 z_1 ^2 }{z_0 ^2 })^{\frac{1}{4}}  }{(1+ 5 \frac{z_1 ^2 }{z_0 ^2 }+ 4 \frac{z_1 ^4 }{z_0 ^4 })^{\frac{1}{4}} } 
\label{muF}
\end{eqnarray}
\begin{figure}[htbp]
\centering
\includegraphics[width=8cm]{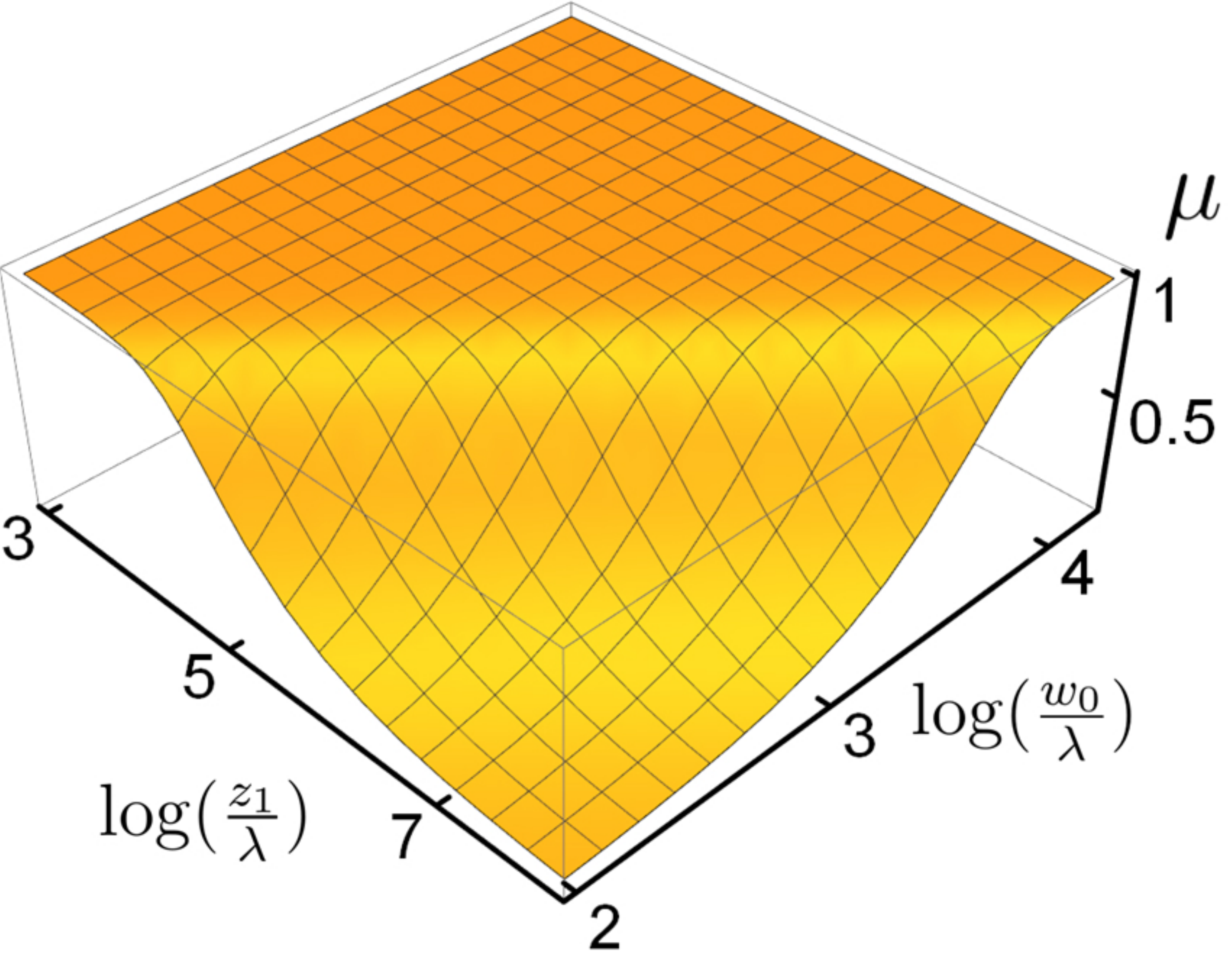}
\caption{ Mode matching value $\mu$ as a function of $w_0$ and $z_1$. }
 \label{modeMu}
\end{figure}

In Fig. \ref{modeMu},  $\mu$ is plotted as  a function of $z_1$ and $w_0$, where $z_1$ is the distance between the mirror and detector 
 We assume that the detector and mirror are large enough that all the waves are detected and reflected. If the distance between the mirror and detector and the size of the beam waist are small enough, the coefficient $\mu$ remains near $1$.

If we consider the case where the vacuum field has waist at the mirror, the coefficient $\mu$ automatically becomes  $1$ due to the symmetry, 
but the vacuum field  $E_v (z_1)$ at the detector does not matche the laser field $E_L (0)$. We assumed that the laser field has beam waist  $w_0$ at the detector, and the vacuum field has a beam waist  $w_m$ at the mirror. 
Then the effective reflectance $\sqrt{R_m} $ becomes

\begin{eqnarray}
\sqrt{R_m} &=&  \frac{|<E_v (z_1)  E_L (0 )^{*} >| }{\sqrt{<E_v (z_1)^2 > <E_L (0 )^2>}}
\nonumber \\
&=&  \frac{\sqrt{2} \sqrt{\frac{w_m }{w_0}}  (1+ \frac{ z_1 ^2 }{z_m ^2 })^{\frac{1}{4}}  }
{(\{(1+ \frac{w_m ^2}{w_0 ^2})^2 + \frac{z_1 ^2 }{z_0 ^2 }\}\{1 +  \frac{z_1 ^2 }{z_m ^2 }\})^{\frac{1}{4}} },
\label{mirR}
\end{eqnarray}
where $z_m = \frac{\pi}{\lambda } w_m ^2$.  

\begin{figure}[htbp]
\centering
\includegraphics[width=8cm]{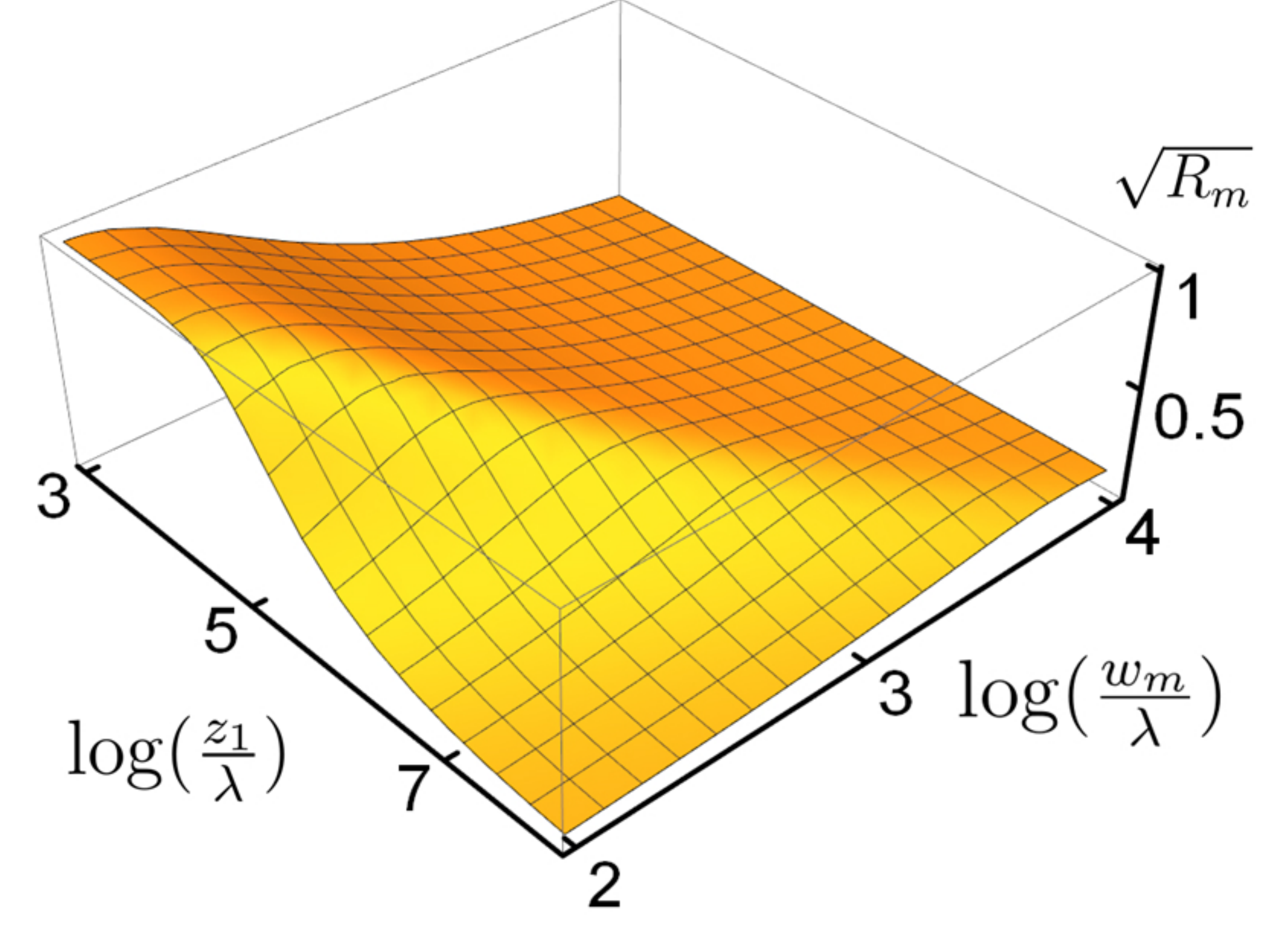}
\caption{Mode matching value $\sqrt{R_m} $ as a function of $w_0$ and $z_1$, with $w_0$ equal to $100 \lambda$  }
 \label{modeR}
\end{figure}
In Fig. \ref{modeR},  $\sqrt{R_m}$ is plotted as a function of $z_1$ and $w_m$, 
where $z_1$ is the distance between the mirror and detector. We set $w_0$ to $100 \lambda$.  Additionally, we also assume that the detector and mirror 
are large enough that all the waves are detected and reflected. The coefficient $\sqrt{R_m}$ can be 1 only when the distance between the mirror and detector is small and the size of the beam waist is sufficiently small. 
 
 The mode matching condition is crucial for detecting the modulation effect of the vacuum fluctuation near the mirror, as denoted by Eq. \ref{flucPr}.  With the usual setup,
  we can not satisfy the conditions $\mu=1$ and $\sqrt{R_m} = 1$. In the next section, we suggest a noble experimental setup that satisfies two mode-matching conditions.

\section{Set up for mode matching}

\begin{figure}[htbp]
\centering
\includegraphics[width=8cm]{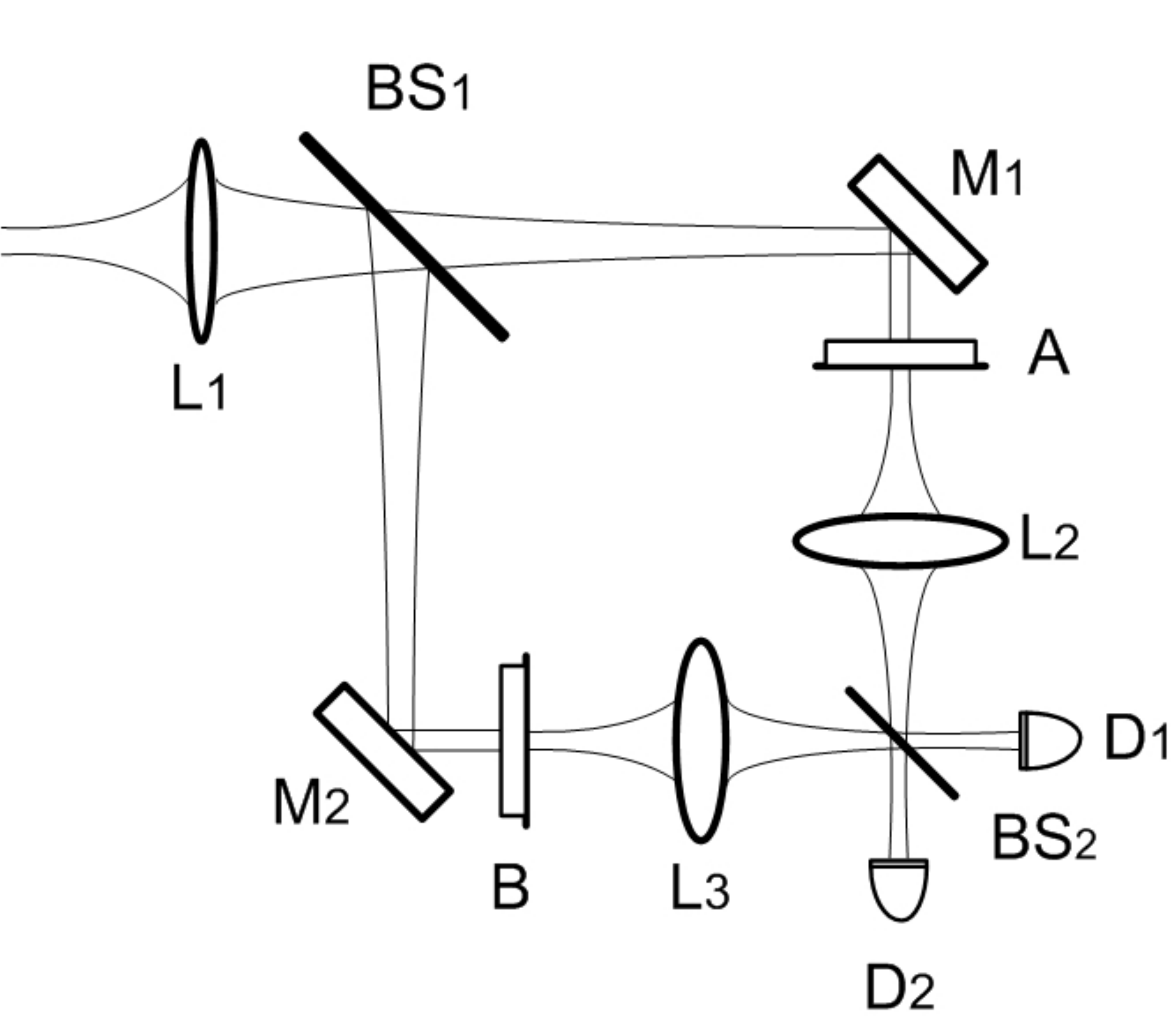}
\caption{Mode matching setup }
 \label{mode}
\end{figure}

For a laser that has a Gaussian transverse mode, we have to establish a vacuum mode that also has a Gaussian transverse mode. Fig. \ref{mode}  displays the setup for perfect mode matching between the laser light mode and a vacuum mode.  

The laser used in the experiment passes through lens $L_1$  and is divided into two by the beam splitter ($BS_1$). 
 The laser is a Gaussian beam and it proceeds according to the Gaussian approximation. The light passing through $BS_1$ and traveling to mirror $M_2$ reaches the partial mirror $B$ and yields a beam waist on the $L_3$ side surface of $B$. Similarly, the light reflecting from the mirror $M_1$ passes through the partial reflector  $A$ and yields a beam waist on the $L_2$ side surface of $A$. 

The light passing through $A$ and $B$ passes through the $L_2$ and $L_3$  of the same focal length, respectively, and yields  another beam waist on the detector surface.
The transmittance of light passing through $A$ from $M_1$  is almost 0, and the reflectance of light stemming from the $L_2$  side is almost 1.
In this way, if the mode is perfectly matched using the light passing through $B$ and $A$, an experimental setup can be established wherein one side of the beam splitter $BS_2$ is a mirror ($A$).

Using this method, the degree of mode matching can be increased compared to that  when the experiment is performed by simply placing a plane mirror on one side of the beam splitter. Additionally the experimental constraints caused by the mode matching can be overcome. 
The experimental setup in Fig. \ref{mode} enables the measurement of how the vacuum fluctuations of the light passing through the beam splitter change when a mirror is placed on one side of the beam splitter.

\section{Conclusion and Discussion.}

The quantum nature of photons is highly dependent on their vacuum fluctuations. Vacuum fluctuations can be directly  measured via homodyne detection. The fluctuation of one quadrature of the vacuum can be less than that of the usual vacuum, e.g., squeezed vacuum. Light intensity fluctuations are also dependent on vacuum fluctuations. Sub-Poisson light can be generated by controlling the vacuum fluctuations based on the nonlinear interaction of light and matter.
In this study, we proposed the modulation of vacuum fluctuations by inserting a mirror on the unused part of the beam splitter in a  homodyne measuring system. 
Furthermore, we calculated the effect of the line width of the laser and the thickness of the detector layer. 
The line width can be practically reduced to modulate vacuum fluctuations, but the decrease of the thickness of the detector to modulate vacuum fluctuations is challenging.
We calculated the effect of  mode matching between the vacuum and light fields and showed that the degree of mode matching obtained by adding a simple mirror 
in the unused beam splitter may not be sufficient to modulate  the vacuum fluctuations.
We present the perfect mode matching method for the vacuum and light fields. Then, the light intensity fluctuations can be reduced by inserting a beam splitter and a mirror. 
We still require a detector with an active layer  thinner than the wavelength to obtain a sub-Poisson light as a function of the distance between the mirror and detector. 
We expect that  our simple method of reducing  vacuum fluctuations will play a great role in quantum information science.



%
%
%

%



%
%

\end{document}